\documentclass{aastex631}
\usepackage{subscript}

\usepackage{xcolor}
\usepackage{soul}
\graphicspath{{./}{figures/}}

\newcommand{\beq}{\begin{equation}}
\newcommand{\eeq}{\end{equation}}

\newcommand{\cm}{cm$^{-2}$}

\newcommand{\Msun}{M$_\odot$}
\newcommand{\kmps}{km~s$^{-1}$}

\newcommand{\Mmol}{{\rm M_{mol}}}
\newcommand{\aco}{\alpha_{\rm CO}}

\newcommand{\hi}{H{\sc i}}

\newcommand{\cii}{[C{\sc ii}]\,158$\mu$m}

\newcommand{\lcou}{K~km~s$^{-1}$~pc$^2$}

\begin{document}

\title{A Massive, Dusty \hi-Absorption-Selected Galaxy at $z \approx 2.46$ Identified in a CO Emission Survey}

\correspondingauthor{Balpreet Kaur}
\email{bkaur@ncra.tifr.res.in}

\author{B. Kaur} 
\affiliation{National Centre for Radio Astrophysics, Tata Institute of Fundamental Research, 
Pune University, Pune 411007, India}

\author{N. Kanekar} 
\affiliation{National Centre for Radio Astrophysics, Tata Institute of Fundamental Research, Pune University, Pune 411007, India}

\author{M. Revalski}
\affiliation{Space Telescope Science Institute, 3700 San Martin Drive, Baltimore, MD 21218, USA}

\author{M. Rafelski}
\affiliation{Space Telescope Science Institute, 3700 San Martin Drive, Baltimore, MD 21218, USA}
\affiliation{Department of Physics \& Astronomy, Johns Hopkins University, Baltimore, MD 21218, USA}

\author{M. Neeleman}
\affiliation{Max-Planck-Institut für Astronomie, Königstuhl 17, D-69117, Heidelberg, Germany}

\author{J. X. Prochaska}
\affiliation{Department of Astronomy \& Astrophysics, UCO/Lick Observatory, University of California, 1156 High Street, Santa Cruz, CA 95064, USA}
\affiliation{Kavli Institute for the Physics and Mathematics of the Universe (Kavli IPMU), 5-1-5 Kashiwanoha, Kashiwa, 277-8583, Japan}

\author{F. Walter}
\affiliation{Max-Planck-Institut für Astronomie, Königstuhl 17, D-69117, Heidelberg, Germany}

\begin{abstract}

We report a NOrthern Extended Millimeter Array (NOEMA) and Atacama Large Millimeter/submillimeter Array (ALMA) search for redshifted CO emission from the  galaxies associated with seven high-metallicity ([M/H]~$\geq -1.03$) damped Lyman-$\alpha$ absorbers (DLAs) at $z\approx1.64-2.51$. Our observations yielded one new detection of CO(3--2) emission from a galaxy at $z=2.4604$ using NOEMA, associated with the $z=2.4628$ DLA towards QSO~B0201+365. Including previous searches, our search results in detection rates of CO emission of $\approx56^{+38}_{-24}$\% and $\approx11^{+26}_{-9}$\%, respectively, in the fields of DLAs with ${\rm [M/H]}>-0.3$ and ${\rm [M/H]}<-0.3$. Further, the \hi-selected galaxies associated with  five DLAs with [M/H]~$>-0.3$ all have high molecular gas masses, $\gtrsim5\times10^{10}\ {\rm M}_\odot$. This indicates that the highest-metallicity DLAs at $z\approx2$ are associated with the most massive galaxies. The newly-identified $z\approx2.4604$ \hi-selected galaxy, DLA0201+365g, has an impact parameter of $\approx7$~kpc to the QSO sightline, and an implied molecular gas mass of $(5.04\pm0.78) \times10^{10}\times(\alpha_{\rm CO}/4.36)\times(r_{31}/0.55)\ {\rm M}_\odot$. Archival Hubble Space Telescope Wide Field and Planetary Camera~2 imaging covering the rest-frame near-ultraviolet (NUV) and far-ultraviolet (FUV) emission from this galaxy yield non-detections of rest-frame NUV and FUV emission, and a $5\sigma$ upper limit of 2.3~M$_\odot$~yr$^{-1}$ on the unobscured star formation rate (SFR).
The low NUV-based SFR estimate, despite the very high molecular gas mass, indicates that DLA0201+365g either is a very dusty galaxy, or has a molecular gas depletion time that is around two orders of magnitude larger than that of star-forming galaxies at similar redshifts.

\end{abstract}


 \section{Introduction} \label{sec:intro}

The highest \hi\ column density absorbers in QSO absorption spectra, the damped Lyman-$\alpha$ absorbers \citep[DLAs, with 
N(\hi)~$\geq 2 \times 10^{20}$~\cm;][]{Wolfe86} are expected to arise in gas associated with galaxies \citep[e.g.][]{Wolfe05}. The presence of a DLA in a quasar spectrum thus allows one to identify high-$z$ galaxies via their Lyman-$\alpha$ absorption signature, rather than via their luminosity. Such \hi-absorption-selected galaxies do not have the luminosity bias that affects emission-selected samples, making them complementary probes of galaxy evolution that are sensitive to galaxies with effectively any mass.

Many attempts have been made, mostly at optical and near-IR wavelengths, to identify and characterize high-$z$ \hi-absorption-selected galaxies, and thereby establish their connection to the emission-selected population \citep[e.g.][]{Warren01,Kulkarni06,Peroux11,Fumagalli15,Krogager17,Mackenzie19}. A number of these studies \citep[e.g.][]{Moller04,Fynbo10,Wang15} have focussed on DLAs with high absorption metallicities, as the correlation between DLA metallicities and the velocity widths of the low-ionization metal lines suggest the existence of a mass-metallicity relation in DLAs \citep[e.g.][]{Ledoux06,Prochaska08,Neeleman13}. However, the presence of the background quasar, which is very bright at these wavelengths relative to the foreground galaxy, has been a significant hindrance to these efforts. As a result, only $\approx 20$~\hi-selected galaxies have been identified via such studies at high redshifts, $z \gtrsim 2$ \citep[e.g.][]{Moller04,Fynbo13,Krogager17,Mackenzie19,Ranjan20}.

In the last few years, with the advent of the Atacama Large Millimeter/submillimeter Array (ALMA), redshifted CO rotational and \cii\ fine-structure emission lines have emerged as a promising tool to identify and characterize \hi-selected galaxies at high redshifts \citep[][]{Neeleman16,Neeleman17,Neeleman19,Peroux19, Klitsch19,Kanekar18,Kanekar20,Kaur21}, yielding a number of surprises.  
For example, \hi-selected galaxies at $z \approx 4$ identified by their \cii\ emission have large impact parameters, $b \approx 16-45$~kpc, from the QSO sightline \citep{Neeleman17, Neeleman19}, far larger than expected for sightlines that pass through the interstellar medium of galaxies. This indicates that the DLA sightlines significantly traverse the circumgalactic medium (CGM), with high \hi\ column densities present in the CGM of high-$z$ galaxies. One of the \cii\ emitters at $z \approx 4.26$ has been shown to be a massive, rotating disk galaxy, the first such cold disk to be identified at $z \gtrsim 4$ \citep{Neeleman20}. 
At $z \approx 2$, ALMA CO searches have found evidence that a number of \hi-selected galaxies associated with high-metallicity DLAs have both remarkably large molecular gas masses, $\gtrsim 5 \times 10^{10} \ {\rm M}_\odot$, and a range of impact parameters, $b \approx 6 - 100$~kpc \citep{Neeleman18,Fynbo18,Kanekar20}. Follow-up multi-line CO studies have found evidence for highly-excited mid-J CO rotational levels in three of the CO-emitting, \hi-selected galaxies at $z \approx 2$ \citep{Klitsch22,Kaur22}.

The very large molecular gas masses of the \hi-selected galaxies at $z \approx 2$ is an especially surprising result of the ALMA CO and \cii\ studies. However, the number of absorber fields at $z \approx 2$ that have been searched for CO emission remains quite small, with 5 CO detections in 12 searches so far \citep{Kanekar20}. The recent upgrade of the NOrthern Extended Millimetre Array (NOEMA) has opened the possibility of carrying out such searches for redshifted CO emission from \hi-selected galaxies at northern declinations, i.e.\ those
inaccessible to ALMA. In this paper, we report a NOEMA and ALMA search for redshifted CO(2--1) or CO(3--2) emission from the fields of seven high-metallicity DLAs at $z \approx 2$, with absorption metallicity [M/H]~$\gtrsim -1.0$, which has resulted in one new CO detection, in the field of the $z \approx 2.4628$ DLA towards QSO~B0201+365. We also report a search for rest-frame far-ultraviolet (FUV) and NUV emission from the new \hi-selected galaxy, using Hubble Space Telescope (HST) Wide Field and Planetary Camera~2 (WFPC2) archival data.

\section{Observations, Data Analysis and Results}
\label{sec:obs}

\subsection{NOEMA and ALMA observations}
\label{sec:obs_CO}

We used NOEMA and ALMA to carry out a search for redshifted CO(2--1) or CO(3--2) emission in the fields of seven DLAs at $z \approx 1.64-2.51$. The targetted DLAs have an absorption metallicity, [M/H] $\geq -1.03$, higher than the median DLA metallicity \citep[$\approx -1.3$; ][]{Rafelski14} in this redshift range. Six of the DLAs were observed with the NOEMA Band-1 receivers between December~2017 and April~2018 (project: W17DG; PI: M. Neeleman), in the D-configuration. The PolyFix correlator was used as the backend, with an instantaneous bandwidth of 8~GHz centred on the redshifted CO(2--1) or CO(3--2) frequency of the target DLA, and sub-divided into 4000~channels, giving a frequency resolution of 2~MHz. The seventh DLA was observed with the ALMA Band-4 receivers (proposal: 2016.1.00628.S; PI: J. X. Prochaska) in January~2017. One of the 1.875~MHz intermediate frequency (IF) bands was used in FDM mode to cover the redshifted CO(3--2) line frequency, with 480~channels and a spectral resolution of 3.9~MHz, while the remaining three IF bands used the TDM mode, with a 2~GHz bandwidth and 128~channels, to cover the continuum emission. The on-source times were $4-9$~hours (NOEMA) and $1.5$~hours (ALMA). We note that the NOEMA and ALMA fields of view (full width at half maximum, FWHM) at the redshifted CO line frequencies are $\gtrsim 50\arcsec$, corresponding to a spatial field of view of $\gtrsim 400$~kpc at the DLA redshifts.\footnote{Throughout the paper, we use a $\Lambda$-CDM cosmology, assuming $\Omega_{\Lambda} = 0.685$, $\Omega_{m} = 0.315$, and H$_0 = 67.4$~\kmps~Mpc$^{-1}$ \citep{Planck2020}.} The observational details are summarized in Table~\ref{tab:obs}.

The initial editing and calibration were carried out in {\sc gildas}\footnote{https://www.iram.fr/IRAMFR/GILDAS} for the NOEMA data, and with the standard ALMA pipeline in the Common Astronomy Software Applications package \citep[{\sc casa} version~5.6;][]{CASA} for the ALMA data. Only one of the seven QSOs, QSO~B0201+365, had a sufficient flux density ($\approx 31$~mJy) for self-calibration; this was carried out in the Astronomical Image Processing System package \citep[``classic'' {\sc aips};][]{Greisen03}, following standard procedures. Any detected continuum emission in each field was then subtracted out from the calibrated spectral-line visibilities, after Fourier-transforming the continuum image to the visibility plane. The residual spectral-line visibilities 
were then imaged in {\sc casa}, using the task {\sc tclean} to produce spectral cubes at velocity resolutions of $25-200$~\kmps. The cubes were then visually inspected to search for redshifted CO(2--1) or CO(3--2) emission.

\begin{figure*}
\centering
\includegraphics [scale=0.14]{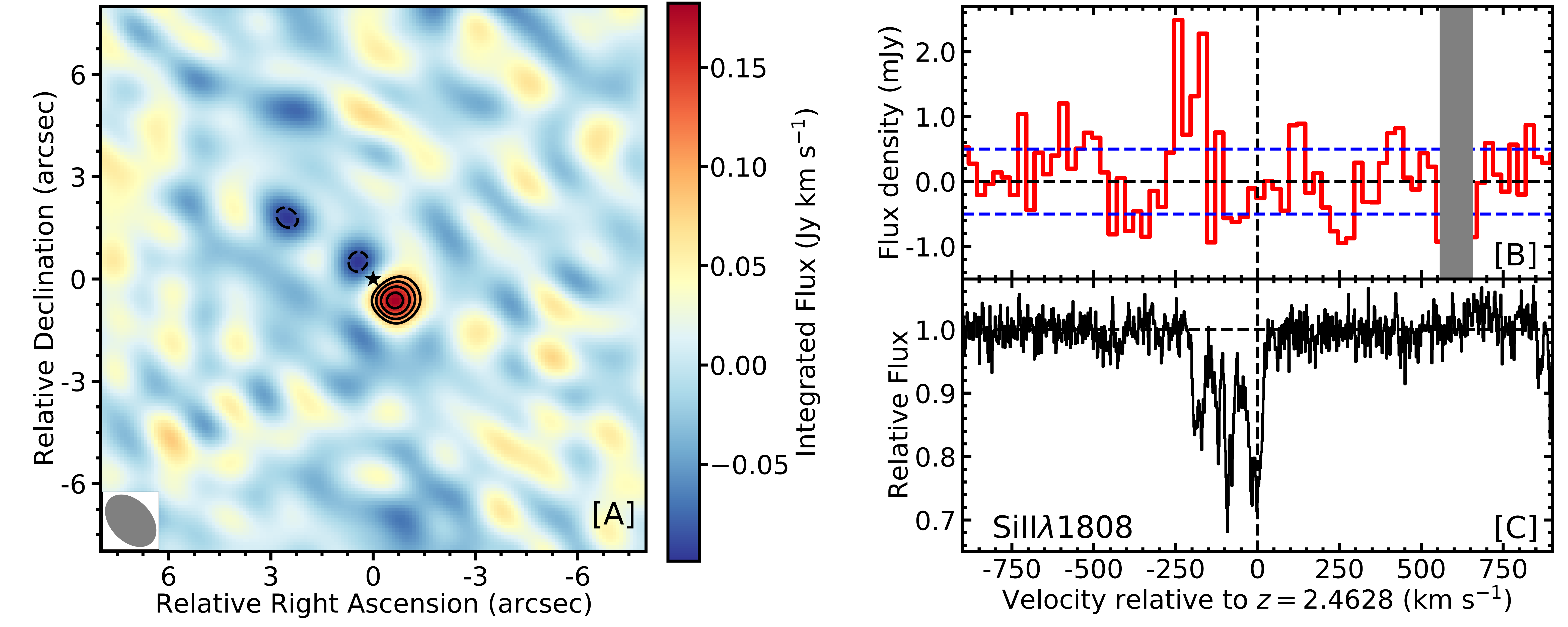}
\caption{[A]~The NOEMA CO(3--2) emission from the \hi-selected galaxy in the field of the $z = 2.4628$ DLA towards QSO~B0201+365. The contours are at $(-4.0, -3.0, 3.0, 4.0, 5.0, 6.0) \times \sigma$ significance, with dashed negative contours; note that there are no $-4\sigma$ contours in the image. The axes are relative to the QSO position, indicated by the black star. We note that the $\approx 3\sigma$ negatives in the image are not aligned with the dirty beam. [B]~The NOEMA CO(3--2) emission spectrum of DLA0201+365g, obtained by taking a cut through the CO cube at the location of the peak in [A].  The blue lines represent the $\pm1\sigma$ error on the spectrum. The vertical grey band indicates channels in the NOEMA frequency coverage that were masked out due to  systematic effects.  These channels have systematically lower visibilities compared to other channels; the origin of this effect is not known. [C]~The Si{\sc ii}$\lambda$1808\AA\ absorption line, from the Keck-HIRES spectrum of \citet{Prochaska96}. In both [B] and [C], the velocity is plotted relative to the redshift of the strongest absorption component, $z=2.4628$.
\label{fig:0201}}
\end{figure*}

We detect CO(3--2) emission from one galaxy (hereafter, DLA0201+365g, at RA=02h04m55.55s, Dec.=+36$^\circ$49$\arcmin$17.3$\arcsec$, $z = 2.4604$) in the field of the $z= 2.4628$ DLA towards QSO~B0201+365. Panels [A] and [B] of Fig.~\ref{fig:0201} show the velocity-integrated CO(3--2) moment-0 image and the CO(3--2) spectrum, the latter obtained from a cut through the peak position of the velocity-integrated image. Both the image and the spectrum show clear detections of the CO(3--2) emission line, at $\approx 6.5\sigma$ statistical significance.  We obtain a velocity-integrated CO(3--2) line flux density of $(0.182 \pm 0.028)$~Jy~\kmps, implying a CO(3--2) line luminosity of $L'_{\rm CO(3-2)} = (6.36 \pm 0.98) \times 10^9$~\lcou. The CO line FWHM is $\approx 125$~\kmps, and the impact parameter to the QSO sightline is $\approx 0.89\arcsec$, i.e. $\approx 7$~kpc at $z = 2.4628$.

No statistically significant ($\geq 5\sigma$) emission was detected in the fields of the remaining six DLAs, at or near the redshifted CO line frequency. The $3\sigma$ upper limits on the velocity-integrated CO(3--2) or CO(2--1) line flux densities are $(0.05-0.14)$~Jy~\kmps, assuming a line FWHM of 200~\kmps, all below the detection
for DLA0201+365g. The corresponding $3 \sigma$ upper limits on the CO(2--1) or CO(3--2) line luminosities are listed in the last column of Table~\ref{tab:obs}.

\subsection{HST-WFPC2 data}

The field of QSO~B0201+365 was observed with HST-WFPC2 in October 1999 with the F450W and F814W filters (program ID: 8085; PI: S. Beckwith). The two filters cover, respectively, the rest-frame FUV and NUV emission from the $z = 2.4604$ \hi-selected galaxy, with two exposures in a single orbit in each filter. After removing cosmic rays with a custom application\footnote{\url{https://github.com/mrevalski/hst\_wfc3\_lacosmic}} of LACOSMIC \citep{vanDokkum01}, the individual exposures in each filter were combined with {\sc DrizzlePac} \citep{hack20}. Specifically, the HST images were aligned with {\sc TweakReg} using the Gaia~DR3 catalogue as the astrometric reference \citep{Gaiadr3}. The routine {\sc AstroDrizzle} was used to drizzle the combined images, with a final pixel fraction of 1 at a pixel scale of 0.0996$\tt ''$/pixel.

Since DLA0201+365g is located close to one of the diffraction spikes of the QSO  point-spread function (PSF), subtraction of the QSO PSF is critical to detect the faint galaxy. The QSO PSF was subtracted using a PSF model obtained from stacking four stars (using a custom code hst\_wfc3\_psf\_modeling\footnote{\url{https://github.com/mrevalski/hst\_wfc3\_psf\_modeling}}) on the same detector chip as the QSO, and separately, from a PSF model obtained by rotating the QSO itself  by 180$^\circ$; similar results were obtained from the two approaches. In the following, we use results from the latter approach.

\begin{figure*}
\centering
\includegraphics [scale=0.17]{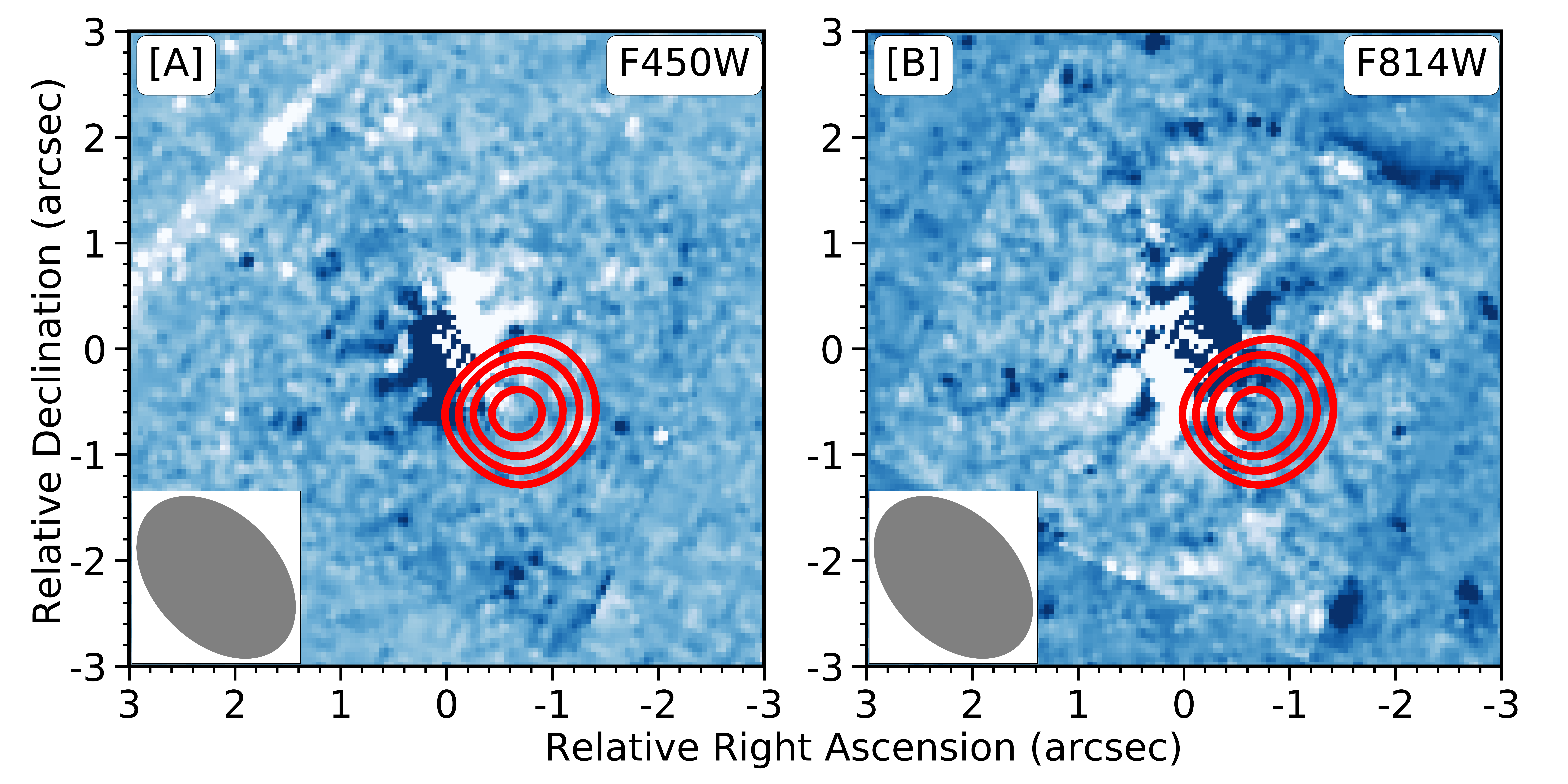}
\caption{The figure shows the HST-WFPC2 [A]~F450W and [B]~F814W images (in colour) of the field of QSO~B0201+365, after subtracting the QSO PSF. The NOEMA velocity-integrated CO(3--2) emission is overlaid (in red contours) on the two HST-WFPC2 images. No emission can be seen in the HST-WFPC2 images at the location of the CO emission. 
\label{fig:0201_hst}}
\end{figure*}

Fig.~\ref{fig:0201_hst} shows overlays of the NOEMA CO(3--2) emission from DLA0201+365g on the HST-WFPC2 F450W and F814W images. We detect no statistically significant emission in the HST images at the location of DLA0201+365g, with $5\sigma$ upper limits of ${\rm m_{AB}} < 24.9$ in the F450W filter and ${\rm m_{AB}} < 25.7$ in the F814W filter, corresponding to the rest-frame FUV and NUV wavelengths for DLA0201+365g, respectively. Assuming a Kroupa initial mass function \citep{Kroupa01} and the SFR calibration of \citet{Kennicutt12}, we obtain $5\sigma$ upper limits on the unobscured SFR of $< 4.9\ {\rm M}_\odot$~yr$^{-1}$ (FUV) and $< 2.3 \ {\rm M}_\odot$~yr$^{-1}$ (NUV).

\begin{table*}
\centering
\caption{Observational details and results. The columns are (1)~the QSO name, (2)~the QSO redshift, (3)~the DLA redshift, (4)~the observed CO rotational transition, (5)~the redshifted CO line frequency, in GHz, (6)~the on-source time, in hours, (7)~the synthesized beam, in arcseconds, (8)~the velocity resolution of the CO spectral cube used for the search, in \kmps, (9)~~the RMS noise on the cube, in $\mu$Jy/Bm, (10)~the integrated CO line flux density, or the $3\sigma$ upper limit to this quantity, assuming a boxcar profile with a width of 200~\kmps, and (11)~the CO(2--1) or CO(3--2) line luminosity, $L{\rm '_{CO}}$, or the $3\sigma$ upper limit to this quantity.
\label{tab:obs}}
\vspace{0.2cm}    
\begin{tabular}{ccccccccccc}
\hline
\hline

QSO        & $z_{\rm QSO}$ & $z_{\rm DLA}$  &  CO line   & $\nu_{\rm obs}$  & Time  &     Beam  & $\delta$V & $\rm RMS_{CO}$ &  $\int S_\mathrm{{CO}} \; \delta V$  &  $L\rm'_{CO}$ \\
           &               &                &       & (GHz)& (hr) &     $''\times''$   & (\kmps) & ($\mu$Jy/Bm)   & (Jy~\kmps) & ($10^{9}$~K~\kmps~pc$^2$)\\
                                                        
\hline                                                  
                                                        
J0927+5823   & 1.91          & 1.6352         &  2--1   &      87.49       & 4.0   &       $2.0 \times 1.6$ & 200 & 173     &    $< 0.11$      & $< 4.0$       \\  
J1049$-$0110$^\ast$ & 2.12   & 1.6577         &  3--2   &     130.11       & 1.5   &       $2.3 \times 1.9$ & 200 & 77      &    $< 0.049$     & $< 0.81$       \\     
J1310+5424   & 1.93          & 1.8006         &  2--1   &      82.32       & 4.1   &       $3.8 \times 2.7$ & 200 & 215     &    $< 0.14$     & $< 5.9$        \\     
J1417+4132   & 2.02          & 1.9509         &  2--1   &      78.13       & 9.0   &       $2.5 \times 1.6$ & 200 & 184     &    $< 0.12$     & $< 5.8 $       \\     
J1555+4800   & 3.30          & 2.3911         &  3--2   &      101.97      & 4.5   &       $2.9 \times 2.3$ & 200 & 198     &    $< 0.13$     & $< 4.0 $       \\     
B0201+365    & 2.91          & 2.4628         &  3--2   &      99.86       & 3.4   &       $1.8 \times 1.2$ & 25  & 500     & $0.182\pm0.028$ & $6.36\pm0.98$  \\               
J1610+4724   & 3.22          & 2.5066         &  3--2   &      98.61       & 7.1   &       $2.3 \times 1.7$ & 200 & 130     &    $< 0.083$     & $< 2.8$         \\   

\hline
\end{tabular}
\vskip 0.1in
$^\ast$J1049-0110 was observed with ALMA, while the remaining six targets were observed with NOEMA.
\end{table*}

\begin{table*}
\centering
\caption{Absorption properties of the seven DLAs, and emission properties of the associated \hi-selected galaxies. The columns are (1)~the QSO name, (2)~the DLA redshift, (3)~the logarithm of the \hi\ column density (in \cm), (4)~the absorption metallicity, [M/H], (5)~the velocity spread of the low-ionization metal lines, $\Delta {\rm V_{90}}$, in \kmps,  (6)~the CO(1--0) line luminosity or the $3\sigma$ upper limit to this quantity, in $10^9$~K~\kmps~pc$^2$, assuming $\rm r_{21} = 0.77$ and $\rm r_{31} = 0.55$ \citep{Tacconi20}, (7)~the molecular gas mass of the \hi-selected galaxy, or the $3\sigma$ upper limit to this quantity, in $10^{10}$~\Msun, and (8)~references for the DLA absorption properties. 
\label{tab:results}}
\vspace{0.2cm}    

\begin{tabular}{ccccccccc}
\hline
\hline
\centering

QSO & $z_{\rm abs}$ & log({N\textsubscript{\hi}}/\cm) &  {[M/H]}     & $\Delta$V$_{90}$                     & $L\rm'_{CO(1-0)}$ 	&  $\Mmol$ 	  & Ref.       \\
	   & 		       & 			               & 	         & \kmps 	                             & $10^{9}$~K~\kmps~pc$^2$ &  $\times 10^{10}$ \Msun  & \\

\hline 

J0927+5823 &   1.6352     &       $20.40 \pm 0.15$    &  $+0.06$    & 208   &  $  <  5.14         $             &       $   <  2.24           $  &     1   &      \\
J1049--0110 &   1.6577     &       $21.35 \pm 0.15$    & $-1.03$    & 330   &  $  <  0.15         $             &       $   <  0.64           $  &    1   &       \\
J1310+5424 &   1.8006     &       $21.45 \pm 0.15$    &  $-0.52$    & 86    &  $  <  7.62         $             &       $   <  3.32           $   &    1   &		\\
J1417+4132 &   1.9509     &       $21.85 \pm 0.15$    &  $-0.93$    & 114   &  $  <  7.52         $             &       $   <  3.28           $ &        1        &		\\
J1555+4800 &   2.3911     &       $21.50 \pm 0.15$    &  $-0.46$    & 199   &  $  <  7.19         $             &       $   <  3.14           $  &      1            &		\\
B0201+365  &   2.4628     &       $20.38 \pm 0.04$    &  $-0.24$    &  200    &  $ 11.6 \pm 1.8    $             &       $   5.04 \pm 0.78     $  &      2, 3            &		\\
J1610+4724 &   2.5066     &       $21.00 \pm 0.15$    &  $-0.35$    & 155   &  $  <  5.12         $             &       $   <  2.23           $  &		  1        &  \\

\hline    
\multicolumn{8}{l}{References: (1) \citet{Berg15}, (2)~\citet{Prochaska96}, (3)~\citet{Neeleman13}.}

\end{tabular}
\vskip 0.1in

\end{table*}

\begin{figure*}
\centering
\includegraphics [scale=0.3]{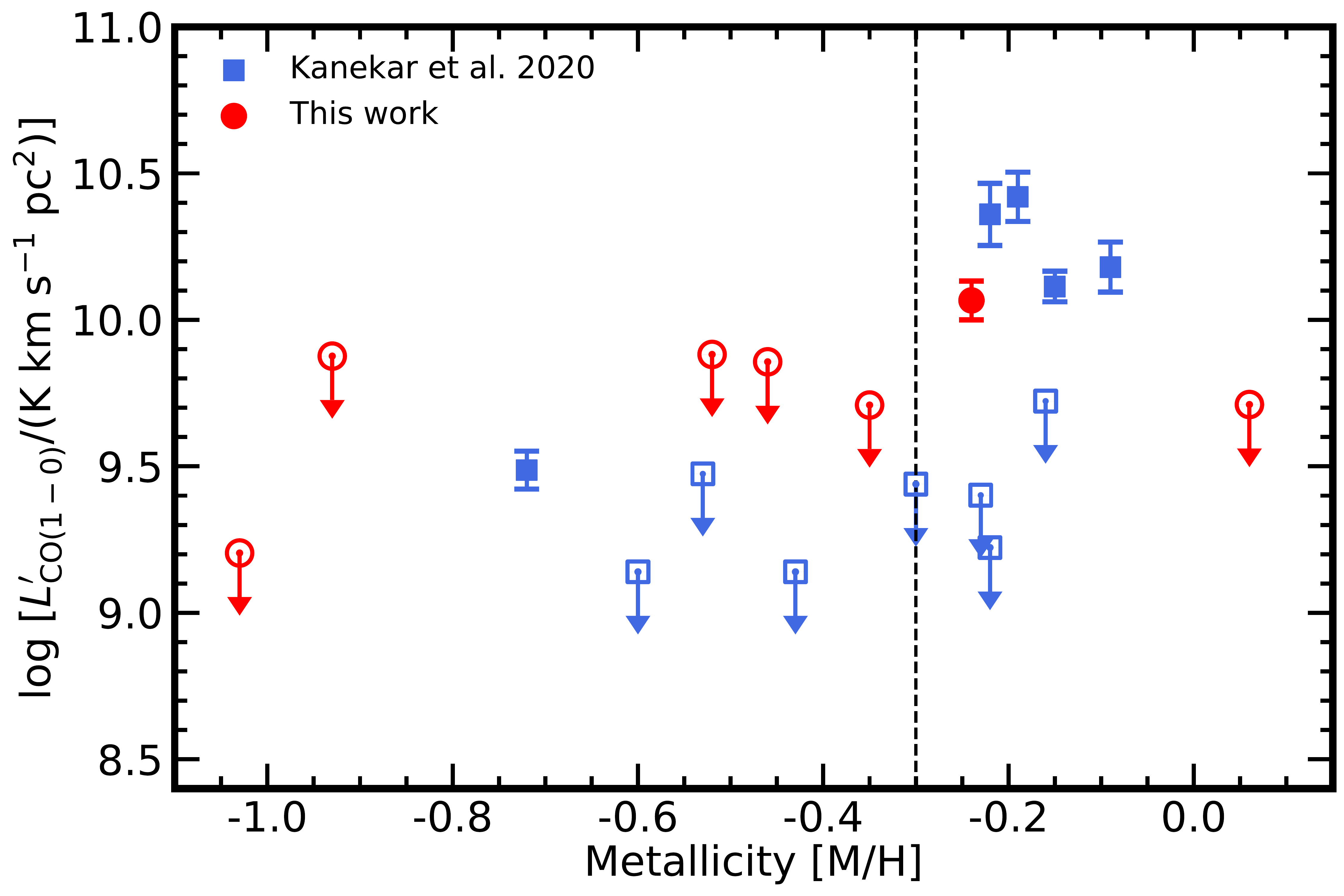}
\caption{The CO(1--0) line luminosity plotted against the DLA absorption metallicity for the nineteen DLAs at $z \approx 2$ with searches for redshifted CO emission. The red circles indicate measurements from the present work, while the blue squares show the sample of \citet{Kanekar20}; CO detections and non-detections are shown, respectively, with solid and open symbols, the latter with downward-pointing arrows. For two DLAs, B1228-113 at $z = 2.1929$ and J0918+1636 at $z = 2.5832$, we show the actual measured CO(1--0) line luminosity \citep{Kaur22}; for the remaining 17 systems, the CO(1--0) line luminosity is inferred from measurements of the CO(2--1), CO(3--2), or CO(4--3) luminosities, assuming sub-thermal excitation with $r_{21}= 0.77$, $r_{31} = 0.55$, and $r_{41} = 0.42$ \citep{Tacconi20}. The dashed vertical line indicates the metallicity [M/H]~$=-0.3$, above which we find a higher CO detection fraction, $\approx 56$\%.
\label{fig:covsmet}}
\end{figure*}

\section{Discussion}

Our new NOEMA and ALMA CO searches have yielded one detection of redshifted CO emission from the \hi-selected galaxy DLA0201+365g in the field of the $z=2.4628$ DLA towards QSO~B0201+365, and six non-detections of CO emission in the fields of the other DLAs. The measured CO line luminosities of Table~\ref{tab:obs} can be converted to molecular gas masses via the relation, $\Mmol = \aco \times {\rm r_{J1}} \times L'_{\rm CO(J \rightarrow J-1)}$, where $\aco$ is the CO-to-H$_2$ conversion factor \citep[e.g.][]{Bolatto13} and ${\rm r_{J1}} \equiv L'_{\rm CO(J \rightarrow J-1)} / L'_{\rm CO(1-0)}$ is the ratio of CO(${\rm J \rightarrow J-1}$) and CO($1-0$) line luminosities \citep[e.g.][]{Carilli13}. The value of of $\aco$ depends on physical conditions in the gas, including temperature, density, and metallicity \citep{Bolatto13, Carilli13, Tacconi20}. Following \citet{Kanekar20}, we assume $\aco = 4.36 \ {\rm M}_\odot$~(K~\kmps~pc$^{2}$)$^{-1}$ for all targets, a reasonable assumption for non-starburst galaxies with near-solar metallicity.  We further assume sub-thermal excitation of the $J=2$ and $J=3$ rotational levels, with ${\rm r_{21}} = 0.77$ and ${\rm r_{31}} = 0.55$, applicable to high-$z$ main-sequence galaxies \citep{Daddi15,Tacconi20}. In the case of DLA0201+365g, we obtain a high molecular gas mass of $\Mmol = (5.04 \pm 0.78) \times ({\rm r_{31}}/0.55) \times (\aco/4.36) \times 10^{10}$~\Msun. For the CO non-detections in the six DLA fields, our upper limits on the CO(2--1) or CO(3--2) line luminosity yield $3\sigma$ upper limits of  $\Mmol < (0.6-3.3) \times 10^{10}$~\Msun\ on the molecular gas mass of the \hi-selected galaxies associated with the DLAs. Table~\ref{tab:results} summarizes the absorption and emission properties of the DLAs and the \hi-selected galaxies.

We note that \citet{Klitsch22} find tentative evidence for a lower $\aco$ value in two of the $z \approx 2$ \hi-selected galaxies of the sample of \citet{Kanekar20}, at $z \approx 2.1929$ towards PKS~B1228-113 and $z \approx 2.5832$ towards QSO~J0918+1636. These two galaxies also show near-thermal excitation of the ${\rm J=3}$ rotational level \citep{Kaur22}. However, these galaxies have the highest measured CO(3--2) line luminosities of the sample, and it is not clear whether their properties are representative of the full sample. Larger samples are needed to explore the excitation conditions of high-$z$ \hi-selected galaxies. 

The newly-detected \hi-selected galaxy DLA0201+365g has a high inferred molecular gas mass, $\approx 5 \times 10^{10} \ {\rm M}_\odot$. Remarkably, the CO line FWHM is only $\approx 125$~\kmps, significantly smaller than the CO FWHMs \citep[$\gtrsim 300$~\kmps; ][]{Kanekar20} of the other \hi-selected galaxies at $z \approx 2$ with CO detections. This suggests that DLA0201+365g has a low inclination to our line of sight. Conversely, the low-ionization metal absorption lines are both wide, $\Delta {\rm V_{90}} \approx 200$~\kmps\ \citep{Prochaska96,Neeleman13}, and asymmetric, with the strongest Si{\sc ii}$\lambda$1808\AA\ absorption of Fig.~\ref{fig:0201}[C] seen to lie at one edge of the line, $\approx 200$~\kmps\ redward of the CO emission. 
 
It is not straightforward to reconcile the narrow observed CO emission with the wide metal absorption lines. One possibility is that the CO emission is detected only from the central regions of the galaxy, which may have higher excitation of the mid-J CO rotational levels, yielding stronger CO(3--2) emission. Alternatively, part of the low-ionization metal line absorption may arise in dense clumps in the CGM of the \hi-selected galaxy, or possibly in a companion galaxy not detected in CO emission.

Our non-detection of DLA0201+365g in the HST-WFPC2 F814W image of Fig.~\ref{fig:0201_hst}, covering its rest-frame NUV emission, implies an upper limit of $2.3 \ {\rm M}_\odot$~yr$^{-1}$ on the 
unobscured SFR of the \hi-selected galaxy. \citet{Wang15} used the Gemini Near-Infrared Integral Field Spectrometer (NIFS) to search for redshifted H$\alpha$ emission from galaxies in the field of the $2.4628$ DLA, reporting a non-detection of H$\alpha$ emission. We re-inspected the Gemini-NIFS H$\alpha$ cube at the spatial position of the CO emission, and found no evidence for H$\alpha$ emission at the position of the CO emission. The H$\alpha$ non-detection places the $3\sigma$ upper limit of $\approx 2.2 \ {\rm M}_\odot$~yr$^{-1}$ on the SFR of DLA0201+365g \citep{Wang15}, similar to the SFR limit from the non-detection of rest-frame NUV emission. We emphasize that these limits are on the dust-unobscured SFR, as any dust obscuration would attenuate the rest-frame NUV and the H$\alpha$ emission.

Combining our molecular gas mass estimate with the upper limit on the rest-frame NUV SFR allows us to estimate the molecular gas depletion timescale, $\rm t_{dep} \equiv \Mmol$/SFR, of DLA0201+365g. This is the timescale on which all the molecular gas of a galaxy would be converted to stars at its current SFR, assuming no replenishment of the gas reservoir.  We obtain $\rm t_{dep} > 22$~Gyr for DLA0201+365g ($> 15$~Gyr, on including the $2\sigma$ uncertainty in the molecular gas mass), more than an order of magnitude larger than the molecular gas depletion timescales in star-forming galaxies at $z \approx 2.5$ \citep[e.g.][]{Tacconi20} or in the local Universe \citep[e.g.][]{Saintonge17}. 

 Large molecular gas depletion timescales ($3-120$~Gyr) have also been found in \hi-selected galaxies at intermediate redshifts, $z \approx 0.5-0.8$ \citep{Moller18,Kanekar18}. However, in at least one case here, the $z \approx 0.7163$ \hi-selected galaxy J1323-0021g, the dust-obscured SFR, estimated from the Herschel image, is $\approx 3$ times higher than the SFR estimate from the H$\alpha$ line \citep{Moller18}. This implies a significantly lower molecular gas depletion time, $\approx 5$~Gyr, when obscuration effects are taken into account.

 Similar to the $z \approx 0.7163$ \hi-selected galaxy, J1323-0021g, a more plausible explanation of the low NUV and H$\alpha$ SFR estimates in DLA0201+365g is dust obscuration of the NUV and H$\alpha$ emission. An independent estimate of the total SFR may be obtained from the typical molecular gas depletion timescales of star-forming galaxies at the DLA redshift. We estimate $\rm t_{dep} \approx 0.22$~Gyr for DLA0201+365g from the median of measurements of $\rm t_{dep}$ in star-forming galaxies at $z \approx 2.46$ \citep{Tacconi20}. Combining this with the molecular gas mass estimate of $5.04 \times 10^{10} \ {\rm M}_\odot$ gives a total SFR of $\approx 230 \ {\rm M}_\odot$~yr$^{-1}$, two orders of magnitude larger than the upper limits on the SFR from the rest-frame NUV and H$\alpha$ non-detections. We note that even a molecular gas depletion time of $\approx 1-2$~Gyr, similar to that of galaxies in the local Universe \citep[e.g.][]{Leroy13, Saintonge17}, would yield a total SFR of $\approx 50 \ {\rm M}_\odot$~yr$^{-1}$. This suggests that DLA0201 is likely to be a very dusty galaxy.

The ratio of the Zn abundance to the Fe (or Cr) abundance in DLAs has been used to quantify the extent of depletion of refractory elements onto dust grains  \citep[e.g.][]{Pettini94,Pettini97,Hou01,Prochaska02,Ledoux03,DeCia16}, although challenges exist such as observations showing that Zn traces S \citep{Fenner04,Rafelski12,RomanDuval22}. Specifically, we note that there are caveats in using [Zn/Fe] as a dust indicator, as the origins of Zn are uncertain and the Zn abundance also depends on the star-formation history of a galaxy \citep[e.g.][]{Fenner04}. Regardless, \citet{RomanDuval22} find that [Zn/Fe] generally traces the depletion of heavy elements in DLAs. For DLA0201+365, [Zn/Fe]~$=+0.58$ \citep{Prochaska07}, indicating that the DLA sightline is dusty, with substantial Fe depletion. This is broadly consistent with the result that DLA0201+365g is a dusty galaxy. The depletion in DLA0201+365 is similar to that in the other $z \approx 2$ DLAs that show CO emission from the associated galaxies \citep{Kanekar20}. Five of the six DLAs have [Zn/Fe] (or [Zn/Cr]) in the range $+0.58$ to $+0.91$. DLA2225+0527 is the only absorber of the sample with a significantly higher depletion, [Zn/Fe]~$= +1.23$; here, the background QSO shows significant dust reddening \citep{Krogager16}.

The sample of $z \approx 2$ DLAs with searches for redshifted CO emission consists of 19 systems, with six CO detections and 13 non-detections \citep[][this work]{Neeleman18,Fynbo18,Kanekar20}, i.e. a CO detection rate of $\approx 32$\%. All 19 DLAs have absorption metallicities $\geq -1.03$, i.e. higher the median DLA metallicity \citep[$\approx -1.3$; ][]{Rafelski14} in this redshift range. Fig.~\ref{fig:covsmet} plots the measured or inferred CO(1--0) line luminosity of the \hi-selected galaxies against the absorption metallicity of the corresponding DLAs. The figure shows that a transition from a high CO detection rate to a low detection rate appears to take place at the metallicity [M/H]~$\approx -0.3$, indicated by the dashed vertical line. Five of the nine DLAs with [M/H]$> -0.3$ have detections of CO emission, i.e., a CO detection rate of $56^{+38}_{-24}$\%, while only one of the nine DLAs with [M/H]~$< -0.3$ shows CO emission, yielding a CO detection rate of $\approx 11^{+26}_{-9}$\%. Thus, although the sample remains small, we find evidence for a higher detection rate of CO emission from galaxies associated with high-metallicity DLAs. 

Further, all five CO emitters associated with the highest-metallicity DLAs, with [M/H]~$> -0.3$, have high CO line luminosities, $L'_{\rm CO(1-0)} > 10^{10}$ \lcou. This suggests that the CO line luminosities of \hi-selected galaxies may depend on the absorption metallicity. We tested this hypothesis by carrying out Peto-Prentice and Gehan generalized Wilcoxon tests, using the {\sc asurv} package \citep{Feigelson85}, comparing the CO line luminosity distributions of galaxies associated with two sub-samples of DLAs, with metallicities above and below [M/H]$ = -0.3$. We find that the null hypothesis, that the two samples are drawn from the same CO line luminosity distribution, is ruled out at $\approx 2.3\sigma$ significance. We thus find evidence that both the CO detection rate and the CO line luminosity are higher in \hi-selected galaxies associated with the highest-metallicity DLAs, with [M/H]~$> -0.3$.

\section{Summary}

We have used NOEMA and ALMA to search for redshifted CO(2--1) or CO(3--2) emission in the fields of seven high-metallicity ([M/H]~$\geq -1.03$) DLAs at $z \approx 1.64-2.51$. Our searches yielded one CO detection and six non-detections of CO emission, the latter providing $3\sigma$ upper limits of $(0.6-3.3) \times 10^{10} \ {\rm M}_\odot$ on the molecular gas masses of the associated galaxies. Thus far, including the work presented here, the fields of 19 high-metallicity DLAs at $z \approx 2$ have been searched for CO emission from the associated \hi-selected galaxies, with 6 detections of CO emission, i.e. a CO detection fraction of $\approx 32\%$. We find evidence for both a higher CO detection fraction and higher CO line luminosities in \hi-selected galaxies associated with the highest-metallicity DLAs, with [M/H]~$> -0.3$.
The new NOEMA detection of redshifted CO(3--2) emission is from DLA0201+365g, at $z=2.4604$, associated with the $z = 2.4628$ DLA towards QSO~B0201+365. We obtain a high molecular gas mass, $\Mmol = (5.04 \pm 0.78) \times 10^{10} \times (\alpha_{\rm CO}/4.36) \times (r_{31}/0.55) \ {\rm M}_\odot$ for this galaxy. The non-detection of rest-frame FUV and NUV emission from DLA0201+365g in HST-WFPC2 images of the field places an upper limit of $\approx 2.3 \ {\rm M}_\odot$~yr$^{-1}$ on its dust-unobscured SFR, similar to the limit from the earlier non-detection of H$\alpha$ emission. This would imply that the galaxy has an extremely large molecular gas depletion timescale, $\gtrsim 22$~Gyr, more than an order of magnitude larger than that of star-forming galaxies at similar redshifts or in the local Universe. More plausibly, if the galaxy has a molecular gas depletion timescale similar to that of other galaxies at $z \approx 2.46$, we obtain a total SFR of $\approx 230 \ {\rm M}_\odot$~yr$^{-1}$, two orders of magnitude higher than the upper limits from the NUV and H$\alpha$ non-detections, implying that the object would be a very dusty galaxy. The CO emission is significantly narrower than the spread of the low-ionization metal absorption lines suggesting that either the CO emission arises from only the central regions of the galaxy, or a significant fraction of the absorption arises in the CGM.

\software{DrizzlePac \citep{hack20}, {\sc lacosmic} \citep{vanDokkum01}, {\sc casa} \citep[v5.6;][]{CASA}, {\sc aips} \citep{Greisen03}, {\sc asurv} \citep{Feigelson85}, {\sc astropy} \citep{astropy13}.}

\begin{acknowledgments}

BK and NK acknowledge the Department of Atomic Energy for funding support, under project 12-R\&D-TFR-5.02-0700. MN acknowledges support from ERC advanced grant 740246 (Cosmic\_Gas). We thank an anonymous referee for comments on an earlier version of the manuscript, Nichol Cunningham for much help with calibrating the NOEMA data, and Wei-Hao Wang for discussions and for providing us with the analysed Gemini-NIFS spectral cube. This work is based on observations carried out under project number W17DG [Sep-17] with the IRAM NOEMA Interferometer. IRAM is supported by INSU/CNRS (France), MPG (Germany) and IGN (Spain). This paper makes use of the following ALMA data: ADS/JAO.ALMA\#2016.1.00628.S. ALMA is a partnership of ESO (representing its member states), NSF (USA) and NINS (Japan), together with NRC (Canada), MOST and ASIAA (Taiwan), and KASI (Republic of Korea), in cooperation with the Republic of Chile. The Joint ALMA Observatory is operated by ESO, AUI/NRAO and NAOJ. This research is based on observations made with the NASA/ESA Hubble Space Telescope obtained from the Space Telescope Science Institute, which is operated by the Association of Universities for Research in Astronomy, Inc., under NASA contract NAS 5–26555. These observations are associated with program~8085. This material is based upon work supported by the National Science Foundation under Grant No. 2107989.

\end{acknowledgments}

\bibliographystyle{aasjournal}
\bibliography{CO}
\end{document}